\documentclass[aps,prd,preprint,tightenlines,floatfix]{revtex4}
\usepackage{graphicx}
\usepackage{wrapfig}
\newcommand{\nc}{\newcommand}

\nc{\beq}{\begin{equation}}
\nc{\eeq}{\end{equation}}
\nc{\bea}{\begin{eqnarray}}
\nc{\eea}{\end{eqnarray}}
\nc{\n}{\nonumber \\}

\begin{document}

\date{August 19, 2010} 
\title{Caustics, cold flows, and annual modulation}
\author{Aravind Natarajan}
\email{anat@andrew.cmu.edu}
\affiliation{McWilliams Center for Cosmology, Carnegie Mellon University, Department of Physics, 5000 Forbes Ave., Pittsburgh PA 15213, USA}

\begin{abstract} 
We discuss the formation of dark matter caustics, and their possible detection by future dark matter experiments. The annual modulation expected in the recoil rate measured by a dark matter detector is discussed. We consider the example of dark matter particles with a Maxwell-Boltzmann velocity distribution modified by a cold stream due to a nearby caustic. It is shown that the effect of the caustic flow is potentially detectable, even when the density enhancement due to the caustic is small. This makes the annual modulation effect an excellent probe of inner caustics. We also show that the phase of the annual modulation at low recoil energies does not constrain the particle mass unless the velocity distribution of particles in the solar neighborhood is known.
\end{abstract}

\maketitle

\section{Introduction}

Caustics of light have been known since ancient times.  The rainbow is a common example of a light caustic that forms when the family of refracted light rays is projected on to the plane of the sky. Another common example is the heart shaped, or nephroid pattern seen on the bottom of a polished coffee cup. Thus caustics of light are regions where the light intensity is very large.  

\begin{figure}[!h]
\begin{center}
\scalebox{0.6}{\includegraphics{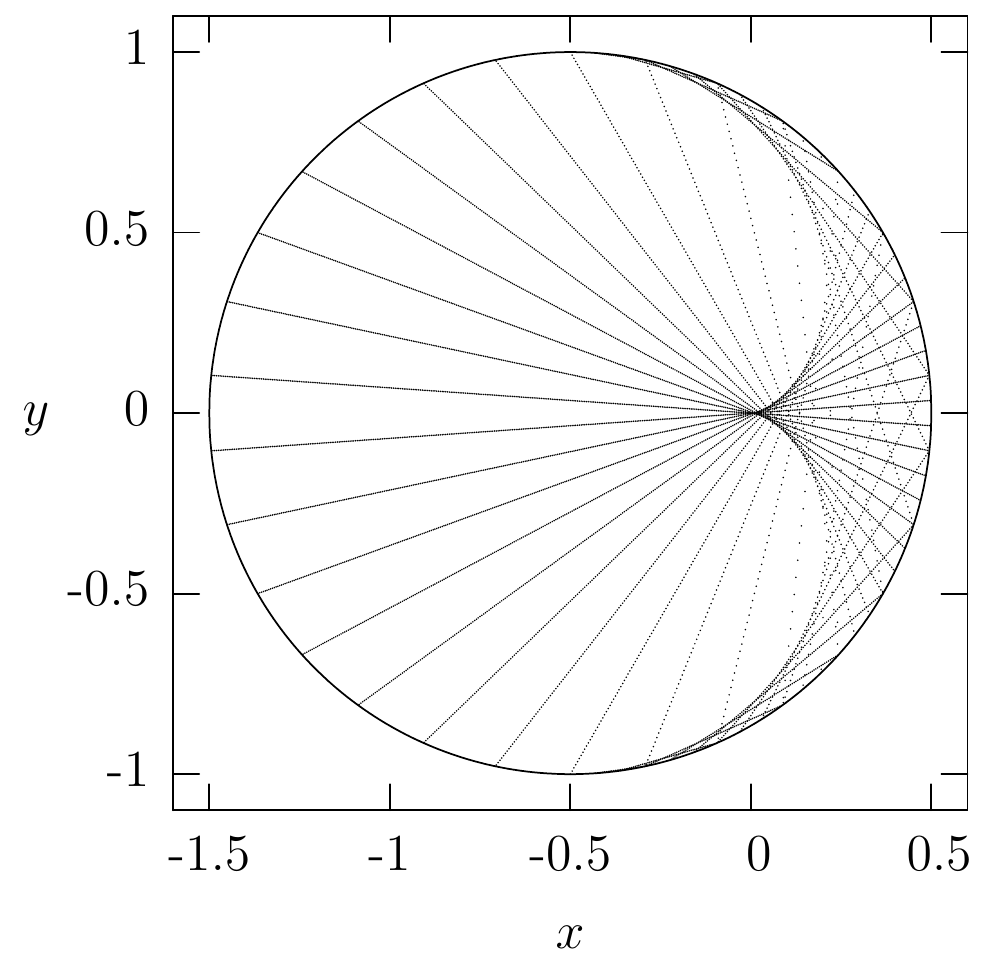}}
\end{center}
\caption{Light rays reflected by a polished metal ring or the inner surface of a reflective coffee cup. The caustic is the envelope of the family of rays.    \label{fig1} }
\end{figure}

Caustics have well defined physical properties. The light intensity on the inner (concave) side of the rainbow varies like the inverse of the square root of distance from the rainbow. However, when the rainbow is approached from the outside (convex) side, the light intensity remains small until the rainbow is reached, at which point it shoots up abruptly. The coffee cup caustic is shown in Fig. 1. The caustic is the envelope of the family of reflected light rays, i.e. the curve tangent to all members of the family of rays. There are two qualitatively distinct regions separated by the caustic: one with three rays at each point and the other with one ray at each point.  Consider the point $(x=0, y=0)$. Near this point, the light intensity varies $\sim |x|^{-1}$ when measured along $x$,  and $\sim |y|^{-2/3}$ when measured along $y$. The intensity at the point $(x=+|\epsilon|,y=0)$ is double the intensity at the point $(x=-|\epsilon|,y=0)$ for small $|\epsilon|$. Near the smooth curve and far from the point $(x=0,y=0)$, the light intensity only varies in the region with 3 rays at each point, like the inverse square root of the distance from the smooth curve. As long as one is far from the boundary between the two regions, a small change in position results in only a small change in light intensity.  However, close to the boundary, a small change in position leads to a very large change in light intensity.

What we have described are called \emph{catastrophes}. Caustics are made up of sections of catastrophes (more precisely, we are concerned with the bifurcation set of the catastrophe). The rainbow caustic is a fold catastrophe. The coffee cup caustic of Fig. 1 consists of 2 fold catastrophes that join together forming a cusp catastrophe. Their properties are described by catastrophe theory which is a branch of singularity theory applied to physical phenomena. Light forms caustics because light is nearly collisionless. Dark matter also has this property. We may thus expect dark matter in galactic halos to form caustics.

Dark matter caustics are singularities in physical space in the limit of zero velocity dispersion \cite{caustic1a,caustic1b,sikivie_ipser}. In reality, the finite velocity dispersion of the dark matter particles cuts off the divergence in the density. Dark matter caustics are then regions of high density.  We may expect a continuous infall of  dark matter to form caustics provided (i) the particles are collisionless and (ii) they have negligible velocity dispersion.

The shells or arcs seen around some giant elliptical galaxies \cite{shells1,shells2} are caustics in the distribution of stars. When a small galaxy is disrupted, the stars of the small galaxy fall into the gravitational potential of the giant galaxy. These stars are then sub-virial, i.e. they are cold. Being compact, stars are nearly collisionless. The infall of cold collisionless stars leads to the formation of caustics at the turnaround radii. The observations of caustics of stars strongly suggests the existence of caustics of dark matter due to the infall of cold, collisionless particles. These are called \emph{outer caustics} and appear as a series of thin spherical shells surrounding galaxies. 

\begin{figure}[!h]
\begin{center}
\scalebox{0.7}{\includegraphics{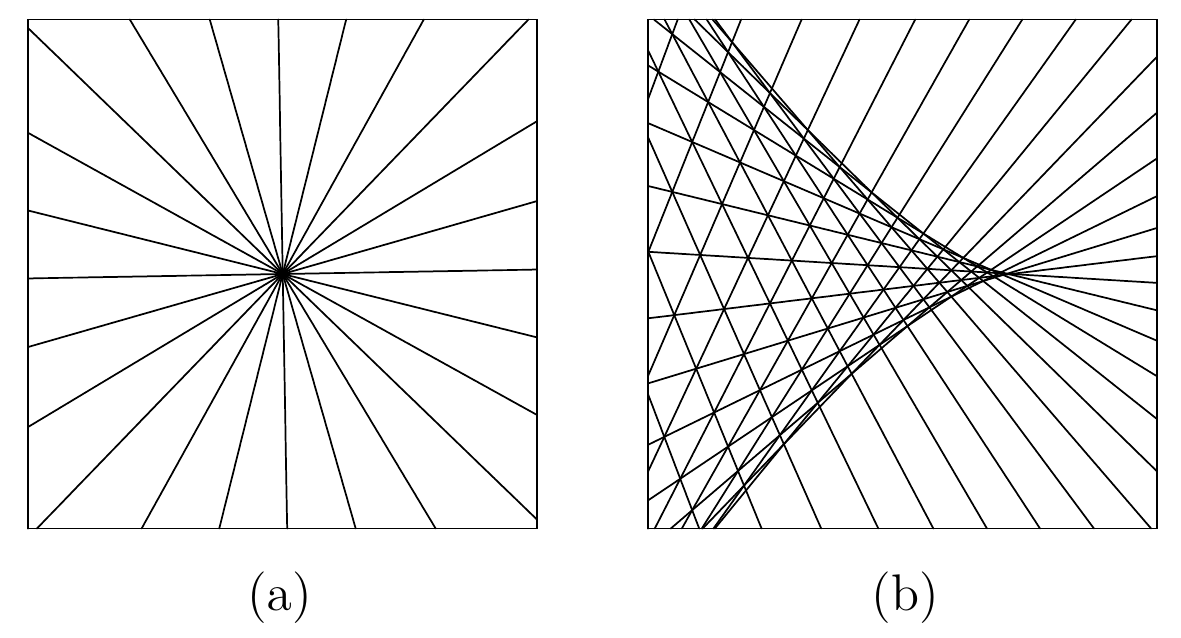}}
\end{center}
\caption{  Dark matter trajectories. (a) shows purely radial orbits, while (b) shows particles with angular momentum.  \label{fig2} }
\end{figure}

Besides the outer caustics, the infall of dark matter particles results in \emph{inner caustics}.  When the particles possess angular momentum, they do not pass through the center, and hence have non-zero inner turnaround radii. Inner caustics form near these locations. Fig. 2 shows the dark matter trajectories for the case of (a) radial infall and (b) infall with angular momentum. When the trajectories are exactly radial, the infall forms a point singularity at the center. When a small amount of angular momentum is added to the flow, the result is a caustic made up of sections of catastrophes. Further small perturbations do not change the qualitative nature of the inner caustic in (b). The inner caustics are made up of folds and cusps which may terminate on higher order catastrophes. The formation of inner caustics is described in detail in \cite{sikivie_crs}. Qualitatively, the caustic of Fig. 2(b) resembles the coffee cup caustic of Fig. 1. The spherical shell caustics mentioned earlier qualitatively resemble the rainbow caustic.

 One may ask whether substructure in real galactic halos can greatly increase the number of flows and weaken the caustics. In  \cite{sikivie_wick}, it was shown that each substructure clump can break up a flow into 3 sub-flows, called ``daughter flows''. Downstream of $p$ clumps, the number of sub-flows $\sim 3^p$ can be very large for moderately large $p$ \cite{arvi1}. However this does not destroy the discreteness of the phase space hypersurface because the density of the main flow is not equally divided among the sub-flows. 

Let us consider the very first flow, i.e. dark matter particles falling towards the halo center for the first time. The first turnaround marks the radius at which particles bound to the halo decouple from the Hubble expansion. A caustic does not form at the first turnaround radius since there is only one flow there. The first outer caustic occurs at the \emph{second} outer turnaround radius. The first inner caustic occurs at the \emph{first} inner turnaround radius. We thus see that the first inner caustic forms before the first   outer caustic. As an example,  let us assume that the first flow contributes $\sim 1\%$ to the local dark matter density in the absence of gravitational scattering by substructure. Let us assume there is enough substructure to break up the first flow into a large number, say $\sim 10^4$ of sub-flows. Let us also (incorrectly) assume that the density is equally divided among the sub-flows. We then have $\sim 10^4$ flows, each contributing $\sim 10^{-4} \%$ to the local dark matter density, and each producing an inner caustic of no realistic importance near the inner turnaround radii. The flows eventually fall out of the potential well (i.e. their radial velocities becomes positive), and form $\sim 10^4$ outer caustics, of negligible density contrast at the outer turnaround radii. The conclusion then would be that there are \emph{no} physically relevant caustics, inner or outer.

This conclusion contradicts the observations of \cite{outer_caustics1, outer_caustics2} who have obtained evidence for the existence of a dark matter outer caustic in their study of galaxies in the group around NGC 5846, and have interpreted the caustic as that of second turnaround, i.e. the first outer caustic. This result has been interpreted by \cite{ring_cluster} as evidence that an inner caustic exists in that group. Indeed it is hard to understand how the first outer caustic can form if the infalling flow is so diffused that it cannot produce the first inner caustic. In \cite{sikivie_ipser, arvi1}, it was found that gravitational scattering of dark matter particles by inhomogeneities does not greatly affect the first few flows unless most of the dark matter is composed of very massive ($ \sim10^{10} M_\odot$) clumps. We know of no evidence that the dark matter in the Milky Way is composed of such massive clumps.  On the contrary, the recent findings of \cite{genel_etal} show that $\approx 40\%$ of the dark matter in halos is accounted for by smooth accretion of particles not bound in halos.

Assuming a continuous infall of cold dark matter particles with angular momentum (assumed aligned with the baryon angular momentum), one may expect to find a series of inner caustics in the plane of the galaxy.     Authors \cite{sikivie_ss1, sikivie_ss2, sikivie_ss3} (motivated by the work of \cite{ss1,ss2})  have proposed a model of the Milky Way halo that predicts the locations of the caustics, and the approximate density near the caustics. In this model, the earth naturally lies between two inner caustics. The distance to the nearest inner caustic has been estimated in \cite{caustics_evidence}. A ring like feature in the cluster Cl 0024+1654  was studied by \cite{ring_cluster} who have interpreted the feature as a dark matter inner caustic.  The presence of inner caustics in spiral galaxies was investigated by \cite{kinney_sikivie}, who found possible evidence for their existence by analyzing the rotation curves of many galaxies. Rises in the rotation curve of the Milky Way also show possible evidence for the existence of caustics in our galaxy \cite{caustics_evidence}. In \cite{caustics_evidence}, the presence of a triangular feature in the infrared spectrum was proposed as possible evidence for the presence of a nearby inner caustic. 

Certain high resolution simulations see cold streams \cite{stiff_widrow} that are associated with inner caustics. However, recent N-body simulations performed by \cite{vogel}  do not find significant inner caustics. The absence of inner caustics in these simulations is inconsistent with the predictions of \cite{arvi1,sikivie_ss1,sikivie_ss2,sikivie_ss3}  and with the claimed observational evidence.  In this article, we do not attempt to reconcile the simulations with the observations. Instead we suggest that a nearby dark matter caustic may be revealed by the velocity distribution of dark matter particles in the solar neighborhood. If there exists a nearby dark matter inner caustic, the dark matter velocity distribution is influenced by the cold flow forming the caustic. We show that the annual modulation expected in the recoil rate measured by a  detector can be used to detect the presence of a nearby inner caustic, even for weak density enhancements.

\section{Formation of caustics}

It was shown in \cite{arvi2} that the continuous infall of cold collisionless matter necessarily produces an inner caustic. Let us review the argument: Consider a sphere with a conveniently chosen radius, such that all particles of a given flow pass through the sphere. Then, we may identify every particle of the flow by 3 parameters $(\tau,\theta,\phi)$. $\tau$ is the time when the particle crossed the sphere on its way in to the halo. $\theta$ and $\phi$ are co-ordinates on the sphere. Let $\vec x = (x,y,z)$ be a location in physical space where the dark matter density is measured. A catastrophe forms at points where the mapping from the space $(\tau,\theta,\phi)$ to the space $(x,y,z)$ is singular. The condition for the mapping to be singular at $(x,y,z)$ is the vanishing of the Jacobian determinant:
\beq
D = det \; \frac {\partial(x,y,z)}{\partial (\tau,\theta,\phi)} = \frac{\partial\vec x}{\partial \tau} \cdot \left( \frac{\partial \vec x}{\partial\theta } \times \frac{\partial \vec x}{\partial\phi} \right ) = 0.
\label{caustic_condition}
\eeq
Let $(\theta_0,\phi_0)$ be the point on the reference sphere where the angular momentum is a maximum (the angular momentum field defined on a sphere has a maximum somewhere unless it is zero everywhere). Let us choose $\vec x_0 (\tau_0, \theta_0, \phi_0)$ to be the point of closest approach of the particles with the most angular momentum, and let us define $r = \sqrt{x^2_0 + y^2_0 + z^2_0}$. Then we have at $\vec x_0$:
\bea
\frac{\partial r}{\partial \tau} = \hat x \cdot \frac{\partial \vec x}{\partial \tau} = 0 \n
\frac{\partial r}{\partial \theta} = \hat x \cdot \frac{\partial \vec x}{\partial \theta} = 0 \n
\frac{\partial r}{\partial \phi} = \hat x \cdot \frac{\partial \vec x}{\partial \phi} = 0. 
\eea
The first condition is due to the reversal of the sign of the radial velocity at $\vec x_0$. The other two conditions are due to the angular momentum maximum at $(\theta_0,\phi_0)$. Since the four vectors $\partial \vec x / \partial \tau$, $\partial \vec x / \partial \theta$, $\partial \vec x / \partial \phi$, and $\hat x$ cannot all be mutually perpendicular at the same point, at least two of the vectors are linearly dependent, or at least one of them is zero. This satisfies the caustic condition Eq. \ref{caustic_condition}. Thus the infall of a cold, collisionless shell of particles necessarily results in the formation of an inner caustic, regardless of assumptions of symmetry.

The structure of the inner caustic depends on the spatial distribution of angular momentum on the initial reference sphere. For purely curl like initial conditions, the inner caustic resembles a ring, made up of sections of the elliptic umbilic catastrophe \cite{sikivie_crs}. For purely gradient like initial conditions, the inner caustic has a more complicated shape and is made up of sections of the hyperbolic umbilic catastrophe \cite{arvi2}. For the special case of axial symmetry with gradient like initial conditions, one may find swallowtail and butterfly catastrophes \cite{arvi2}.

Dark matter caustics can affect the distribution of stars in the galaxy. In \cite{arvi3}, it was proposed that the Monoceros ring of stars may have formed due to the the presence of the $n=2$ inner caustic of the Milky Way. Two mechanisms were found by which the star density could be enhanced in the vicinity of an inner caustic, and these may have played a role in the formation of the Monoceros ring. Authors \cite{lensing1, lensing2, lensing3, lensing4} examined the possibility of observing caustics by gravitational lensing, while \cite{indirect1, indirect2, indirect3, indirect4, indirect5, indirect6, indirect7} have studied dark matter annihilation in caustics. These possibilities however require high densities in the vicinity of the caustic.

Let us now review the annual modulation expected in the recoil rate of a detector, and show that it is an excellent probe of a nearby dark matter caustic even when the density enhancement due to caustic is low. It was shown by \cite{dir_det1, dir_det2} that due to the motion of the earth about the sun, the flux of dark matter particles reaching the earth is modulated with a period of 1 year, and is largest at certain times of the year. DAMA is a dark matter direct detection experiment that measures the annual modulation in the recoil rate, and claims a positive result with $8.9 \sigma$ confidence \cite{dama}. Unfortunately, it seems the results obtained by DAMA are incompatible with the null results of other experiments (see for example \cite{fairbairn, null1, null2}).

The idea of using the annual modulation as a probe of a nearby dark matter caustic is not new. Authors  \cite{caustics_dd1, caustics_dd2, caustics_dd3, gel, ling} have calculated the expected recoil rate for the late infall self similar caustic model. The effect of cold streams has been studied by several authors (see for example \cite{streams1,streams2,streams3,streams4,sav}). Here we review the known results, and study the case of a Maxwellian velocity distribution modified by a single cold flow due to the presence of a nearby caustic. We present results for two cases (i) when the stream contributes significantly to the dark matter density (i.e. caustic with a large density boost) and (ii) when the stream is a small perturbation. It is shown that even in case (ii), the effects are potentially observable.

\section{Direct detection of dark matter.}

Let us consider Weakly Interacting Massive Particles (WIMPs) as the dark matter. The number of recoils per unit time per unit nuclear mass is given by
\beq
dR = \frac{1}{m_{\rm N}} \, \frac{\rho_\chi}{m_\chi} \, \frac{d\sigma}{dQ} dQ \; \left[ v f(v) dv \right ],
\eeq
where $m_{\rm N} = A  m_{\rm p}$ is the mass of the nucleus of the detector, $A$ is the atomic mass number, $m_{\rm p}$ is the nucleon mass, $\rho_\chi$ is the dark matter density at the location of the detector, $\sigma$ is the scattering cross section, $Q$ is the recoil energy, and $v$ is the velocity of the dark matter particles with respect to the detector. For elastic scattering, the energy transferred to the nucleus is given by
\beq
Q = \frac{m^2_{\rm R} v^2}{m_{\rm N}} (1 - \cos\theta),
\label{Q}
\eeq
where $m_{\rm R} = m_\chi m_{\rm N} / (m_\chi + m_{\rm N})$ is the WIMP-nuleus reduced mass and $\theta$ is the scattering angle in the center-of-momentum frame. From Eq. \ref{Q}, we see that in order to observe a nuclear recoil with energy $Q$, the WIMPs must posses a \emph{minimum} speed 
\beq
v_{\rm min} = \sqrt{ \frac{Q m_{\rm N}}{2 m^2_{\rm R}}  },
\label{vmin}
\eeq
since $0 \le (1 - \cos\theta) \le 2$. It is possible for WIMPs to have a larger speed than the minimum given in Eq. \ref{vmin}, but they cannot have a smaller speed for a given recoil energy $Q$.  Here, we consider only the spin-independent cross section. The differential cross section $d\sigma/dQ$ is expressed in the form:
\beq
\frac{d\sigma}{dQ} = \frac{\sigma_0}{Q_{\rm max}} F^2(Q) = \frac{\sigma_0 m_{\rm N}}{2 m^2_{\rm R} v^2} \, F^2(Q).
\eeq
$F(Q)$ is the nuclear form factor. It is obtained from nuclear physics experiments \cite{form_factor1, form_factor2, form_factor3}, and contains the momentum dependence of the cross section:
\beq
F(Q) = \frac{3 j_1(qr)}{ qr } \; e^{-\frac{1}{2} (qs)^2},
\eeq
in units where $\hbar$ and $c$ are set to 1. $q = \sqrt{2Qm_{\rm N}}$, $s=1$ fm, $R = 1.2 A^{1/3}$ fm, $r = \sqrt{R^2 - 5s^2}$, and $j_1$ is the spherical bessel function. It is easy to see that $F^2(Q) \approx 1$ for small $Q$, and falls off at large $Q$.

We therefore have the number of recoils per unit energy, per unit time, per unit detector mass:
\beq
\frac{dR}{dQ} = \frac{\sigma_0 F^2(Q)}{2 m^2_{\rm R} m_\chi} \rho_\chi \, \int_{v_{\rm min}(Q)}^\infty dv \frac{f(v)}{v}.
\eeq 
Let us define the two quantities:
\bea
T(Q,t) &=& \int_{v_{\rm min}(Q)}^\infty dv \frac{f(v)}{v} \n
M(Q_1,Q_2,t) &=& \int_{Q_1}^{Q_2} dQ \, F^2(Q) \, T(Q,t).
\eea
$T(Q,t)$ is called the mean inverse speed and $M(Q_1, Q_2, t)$ is the number of recoils per unit time, per unit detector mass, between energies $Q_1$ and $Q_2$.

Let us now focus our attention on the distribution of WIMP velocities. The spectrum of dark mater velocities is discrete, i.e. a sum over flows:
\beq
f(\vec v) = \sum_i \delta(\vec v - \vec v_i),
\label{flows}
\eeq
where $\vec v_i$ is the velocity of the $i^{\rm th}$ flow. There are a large number of flows that occupy the inner region of phase space, while the fast flows which are few and well separated in velocity, are mostly found in the outer regions of phase space. A detector with finite resolution will likely not be able to resolve the numerous flows in inner phase space. It is useful then to break up the sum over all flows into 2 parts: a sum over cold flows, and a sum over thermal flows. We may replace the sum over thermal flows by a Maxwell-Boltzmann velocity distribution:
\beq
f_{\rm max}(\vec v_{wh}) = \frac{\exp \left[ - (\vec v_{wh} / v_0)^2 \right ]}{\pi^{3/2} \; v^3_0 },
\label{max}
\eeq
where we have ignored the effect of the finite escape velocity. The subscript $wh$ stands for ``WIMP-halo'' and indicates that the velocities are measured in the rest frame of the halo, not the detector. We take $v_0$ to be 220 km/s for the Milky Way. Eq. \ref{flows} can be expressed as:
\beq
f(\vec v) = \sum_{ {\rm cold} \; {\rm flows} \; i } \delta(\vec v - \vec v_i) + f_{\rm max}(\vec v) = f_{\rm cold}(\vec v) + f_{\rm max}(\vec v).
\eeq

Let us now compute the mean inverse speed $T(Q,t)$  for the two distributions. Consider a single cold flow: $f(\vec v_{wh}) = \delta(\vec v_{wh} - \vec v_{fh})$, where the subscript $fh$ stands for ``flow-halo'' and indicates that the flow velocity $v_{fh}$ is measured relative to the halo. Transforming to the earth's rest frame using $\vec v_{fh} = \vec v_{fe} + \vec v_\oplus(t) + \vec v_\odot$, we obtain $f(\vec v_{we}) = \delta(\vec v_{we} - \left[ \vec v_{\rm f \odot} - \vec v_\oplus(t) \right ])$.  $\vec v_{\rm f \odot}$ is the flow velocity relative to the sun, and $\vec v_\oplus(t)$ is the velocity of the earth relative to the sun. The speed of the flow relative to the earth is given by
\bea
|\vec v_{fe}(t)| = |\vec v_{\rm f \odot} - \vec v_\oplus(t)| &=& \left [ v^2_{f \odot} + v^2_\oplus (t) - 2 \vec v_\oplus (t) \cdot \vec v_{f \odot}  \right ]^{1/2}  \n
&\approx&  v_{f \odot} \left [ 1 - \frac{v_\oplus}{v_{f \odot}} \;  \hat v_\oplus (t) \cdot \hat v_{f \odot} \right ],
\label{flow_earth}
\eea
for $v_\oplus \ll v_{f\odot}$. The mean inverse speed is:
\bea
T_{\rm flow}(Q,t) &=& \frac{1}{|\vec v_{fe}(t)|} \; \theta [ |\vec v_{fe}(t)| - v_{\rm min}(Q) ] \n
 &=& \frac{1}{v_{f \odot}} \; \left[ 1 + \frac{v_\oplus}{v_{f \odot}} \; \hat v_\oplus (t) \cdot \hat v_{f \odot} \right ] \; \theta \left[ v_{f \odot} - v_\oplus \, (\hat v_\oplus(t) \cdot \hat v_{f\odot}) - v_{\rm min}(Q) \right ].
\label{T_flow}
\eea
$\theta(v - v_0)$ is the unit step function = 1 for $v > v_0$ and 0 otherwise. $v_\oplus$ is the time averaged value of the earth's velocity about the sun.

Let us now consider the Maxwellian distribution given by Eq. \ref{max}. We again transform to the rest frame of the detector using the relation $\vec v_{wh} = \vec v_{we} + \vec v_{eh} = \vec v_{we} + \vec v_\odot + \vec v_\oplus(t)$ to obtain the speed (1-dimensional) distribution  $f(v_{eh}) = f(v)$:
\beq
f(v) = \frac{v}{\sqrt{\pi} v_0 v_{eh}} \; \left[ e^{-\left( \frac{v-v_{eh}}{v_0} \right )^2 } - e^{-\left( \frac{v+v_{eh}}{v_0} \right )^2 } \right ],
\eeq
giving a mean inverse speed
\beq
T_{\rm max}(Q,t) = \frac{1}{2 v_{eh}(t)} \; \left[ \textrm{erf} \left \{ \frac{v_{\rm min}(Q) + v_{eh}(t)}{v_0} \right \} - \textrm{erf} \left \{ \frac{v_{\rm min}(Q) - v_{eh}(t)}{v_0} \right \} \right ],
\label{Tmax}
\eeq
with $v_{eh}(t) = | \vec v_\odot + \vec v_\oplus(t) |$ given by
\bea
v_{eh}(t) &=& \left [ v^2_\odot + v^2_\oplus (t) + 2 \vec v_\oplus (t) \cdot \vec v_\odot   \right ]^{1/2}  \n
&\approx&  v_\odot \left [ 1 + \frac{v_\oplus}{v_\odot} \;  \hat v_\oplus (t) \cdot \hat v_\odot \right ].
\eea

\subsection{Flow near a caustic.}

As we mentioned in the Introduction, the continuous infall of dark matter results in a series of outer and inner caustics. If the angular momentum of the dark matter is aligned with that of the baryons, these caustics can be expected to lie in the galactic plane. As a result, the earth lies between two inner caustics. In the self-similar infall model of \cite{sikivie_ss1, sikivie_ss2, sikivie_ss3}, the earth is situated close to the fifth inner caustic. The velocity distribution at the earth's location is therefore influenced by the particles forming the caustic.

Let us choose a co-ordinate system in which the $+\hat x$ axis points towards the galactic center, the $+\hat y$ axis points in the direction of galactic rotation, and the $+\hat z$ axis points towards the north galactic pole.  \cite{ling} have listed the velocities of the different dark matter flows (note that the system of co-ordinates used here is different from that used in \cite{ling}). Here, we will consider only the two flows producing the caustic closest to the sun, which have velocities relative to the halo \cite{ling}  $(\pm 100 \, \hat x + 470 \, \hat y +  0 \, \hat z)$ km/s. Relative to the sun, these flows have velocities:
\bea
v_{a \odot} &=& 253.6 \, \textrm{km/s} \; \left( 0.3549 \, \hat x + 0.9345 \, \hat y - 0.0276 \, \hat z \right ) \n
v_{b \odot} &=& 261.4 \, \textrm{km/s} \; \left( -0.4208 \, \hat x + 0.9067 \, \hat y - 0.0268 \, \hat z \right ).
\label{flow_vel}
\eea
\begin{figure}[!h]
\begin{center}
\scalebox{0.7}{\includegraphics{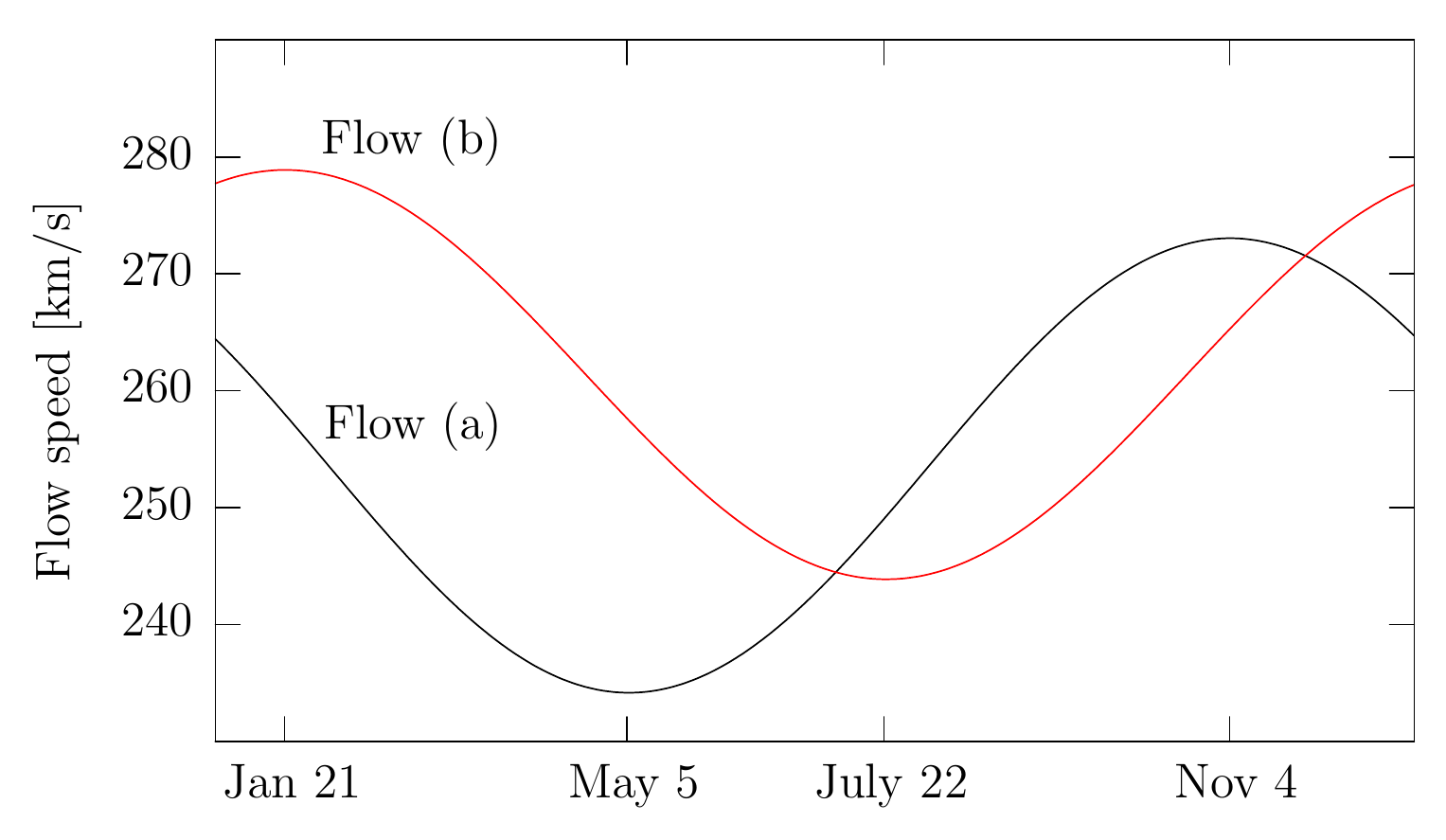}}
\end{center}
\caption{ Speed of the two caustic flows relative to the earth. Flow (a) peaks in November, and is smallest in May. Flow (b) peaks in January, and is smallest in July. \label{fig3} }
\end{figure}

The velocities of the sun (about the halo center) and the earth (about the sun) in these co-ordinates are respectively (see for example \cite{gel,streams2,sav}, and references therein):
\bea
\vec v_\odot &=& 233.3 \, \textrm{km/s} \; \left[ 0.0429 \, \hat x + 0.9986 \, \hat y + 0.0300 \, \hat z \right ] \n
\vec v_\oplus (t) &=& 29.8 \, \textrm{km/s} \; \left [ (0.9931 \cos\phi - 0.0670 \sin\phi) \, \hat x \right. \n
&+& (\left. 0.1170 \cos\phi + 0.4927 \sin\phi) \, \hat y - (0.0103 \cos\phi + 0.8676 \sin\phi) \, \hat z \right ], 
\label{earth}
\eea
where the angle $\phi(t) = 2 \pi \; (t - \textrm{March} \; 21)/365$. We note that earth's velocity about the sun is most closely aligned with the sun's velocity about the halo center when $\phi = 71 ^\circ$, which occurs around June 1. The two velocity vectors are most misaligned when $\phi = 251 ^\circ$ which occurs six months later, around Nov 30.  Fig. \ref{fig3} shows the two caustic flow velocities $v_{a\oplus}(t)$, and $v_{b\oplus}(t)$ relative to the earth.  Using Eq. \ref{flow_earth}, \ref{flow_vel}, and \ref{earth}, we see that $v_{a\oplus}(t)$ is largest when $\phi = 225^\circ$ (around Nov 4) and smallest when $\phi = 45 ^\circ$ (around May 5). $v_{b\oplus}(t)$ is largest when $\phi = 302^\circ$ (around Jan 21) and smallest when $\phi = 122 ^\circ$ (around July 22). 

One of these flows is the dominant flow, and for definiteness, let us assume flow $v_a$ is the dominant flow. This flow shows an $\approx 8.3 \%$ speed modulation with a maximum of $\approx 275$ km/s on $\sim$ November 4, and a minimum of $\approx 232$ km/s on $\sim$ May 5.

Let $\xi$ be the fraction contributed by the dominant caustic flow to the total dark matter density at the earth's location. Then,
\beq
\rho f(v) = \rho \left[ \xi f_{\rm flow}(v) + (1-\xi) f_{\rm max}(v) \right ].
\eeq
$\xi$ can be large if the density enhancement due to the caustic is large. \cite{caustics_evidence, ling} have estimated that the dominant caustic flow, dubbed the ``big flow'' contributes $\sim 73\%$ of the dark matter density at the earth's location. We now show that even if the dominant flow contributes as little as 5\% to the local dark matter density, it has potentially observable consequences.

\begin{figure}[!h]
\begin{center}
\scalebox{0.8}{\includegraphics{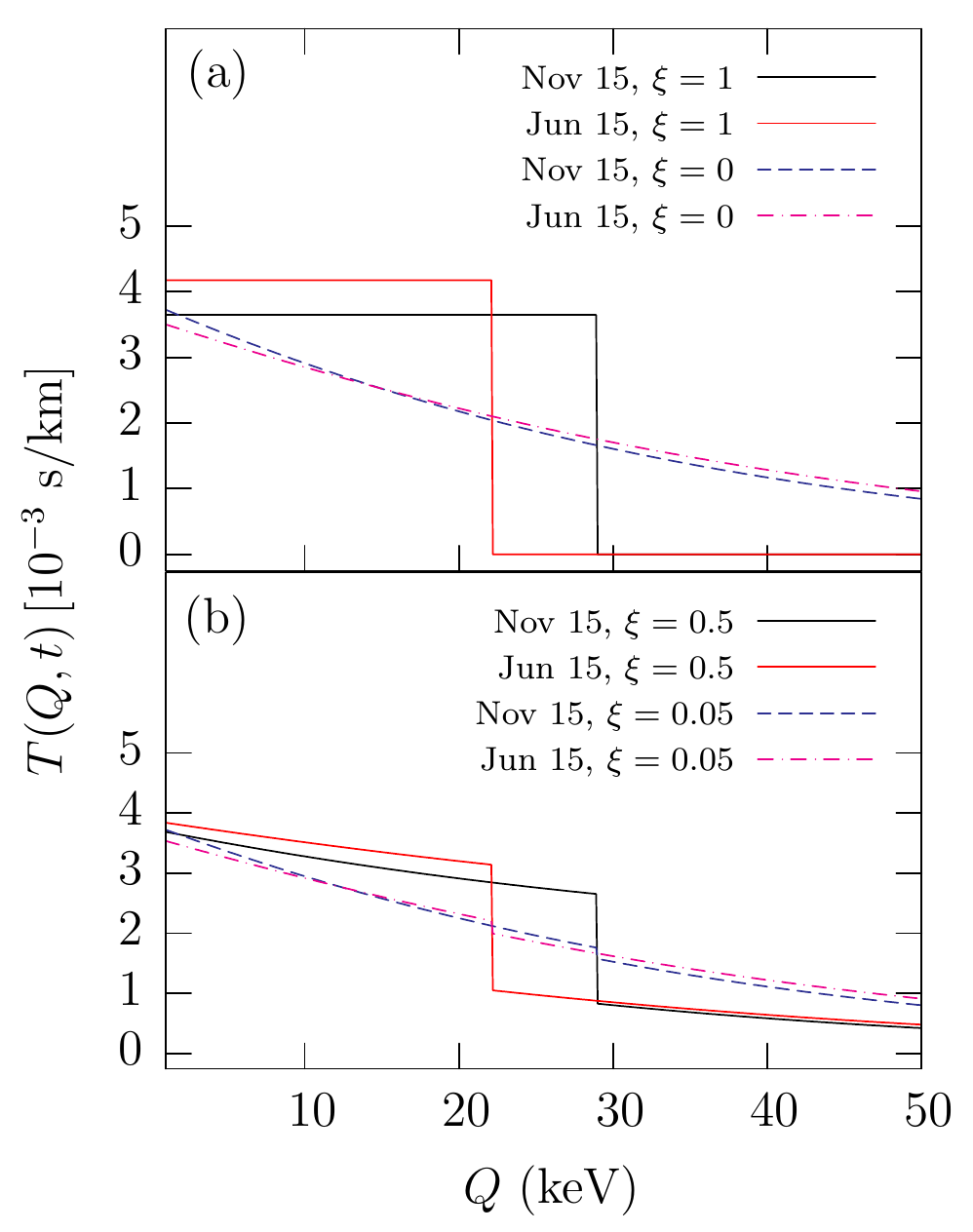}}
\end{center}
\caption{ Mean inverse speed $T(Q,t)$. Panel (a) shows the variation for the caustic flow (a) and for the Maxwellian. Panel (b) shows a combination of the two. Shown are measurements in November and June.   \label{fig4} }
\end{figure}
 
 \subsection{Annual modulation.}
 
Fig. \ref{fig4} shows the mean inverse speed $T(Q,t)$ for an assumed WIMP mass $m_\chi = 70$ GeV, and a Germanium detector with atomic mass number $A = 73$, as a function of recoil energy $Q$. The top panel (a) shows $T(Q,t)$ for the two extreme cases $\xi = 1$ (density entirely due to the caustic flow, shown by the two solid curves black and red) and $\xi = 0$ (caustic flow absent, shown by the dashed curves blue and pink), expected in November, and in June respectively.  For a flow velocity relative to the earth $v_{\rm F}$, the maximum energy at which recoils are measured is given by
\beq
Q_{\rm max} = \frac{2 m^2_{\rm R} v^2_{\rm F} }{m_{\rm N}} \approx 25 \, \textrm{keV} \; \left[ 1 \pm 0.166 \right ],
\eeq
for the assumed values of $m_\chi$, $m_{\rm N}$, and $v_{\rm F}$, with the maximum and minimum values occurring in November and May respectively. The height of the step is given by Eq. \ref{T_flow} and peaks when the speed is the lowest i.e. in May, and is smallest in November when the speed is greatest. Note that the modulation of the edge of the step ($\propto$ square of the flow speed) is twice the modulation of the step height (inverse of the flow speed), and has opposite phase. The flow ceases to be visible to the detector during part of the year starting from $Q \sim 21$ keV (for the assumed values of $m_\chi$ and $m_{\rm N}$). As the energy is increased, $T$ is non-zero close to November, resulting in a maximum then.

Let us look at the behavior of the recoil rate integrated over all energies $M(0,\infty,t)$:
\bea
M(0,\infty,t) &=& \int_0^\infty dQ F^2(Q) \, T(Q,t) = \int_0^{Q_{\rm max}} dQ \frac{F^2(Q)}{v_{f\odot}} \n
         &\sim& \frac{Q_{\rm max}}{v_{f\odot}} \sim v_{f\odot},
\eea
where we ignored the energy dependence of $F^2(Q)$. Thus, when integrated over all recoil energies, the recoil rate peaks in November for this particular flow, and is smallest in June. However, when $M$ is measured over only small recoil energies, we have:
\beq
M(0,Q_{\rm small},t) = \int_0^{Q_{\rm small}} dQ F^2(Q) \, T(Q,t) \sim \frac{Q_{\rm small}}{v_{f\odot}},
\eeq
which implies that the integrated recoil rate peaks in June for small recoil energies $Q$. As $Q$ is increased, the phase abruptly reverses, peaking in November. The abrupt shift in $T$ is because the flow is no longer visible to the detector around June at these energies, owing to the smaller speed around June. In contrast, for a Maxwellian velocity distribution, the recoil rate peaks in November for small $Q$, reverses phase as $Q$ is increased, and peaks in June for sufficiently large $Q$ \cite{mass_phase}.

The two dashed curves show the modulation of $T$ for the purely Maxwellian case. The difference is far less dramatic than for the case of the pure caustic flow. At low recoil energies, $T$ peaks in November (blue dashed curve) and is smallest in June (pink dot-dashed curve). The phase reverses at $\sim$ 15 keV (for the assumed values of $m_\chi=70$ GeV and $A = 73$), and $T$ then peaks in June and has a minimum in November. At high energies, $T$ is small since (i) there are few particles with sufficient energy to effect recoils and (ii) $F^2(Q)$ is small for large $Q$.

Fig. \ref{fig4}(b) shows the more physical case with both the caustic flow and the thermal halo contributing to the recoil energy spectrum. Shown are the two cases $\xi = 0.5$ (strong density enhancement) and $\xi = 0.05$ (weak density enhancement). Let us first consider the case $\xi = 0.5$ (shown by the solid red and black curves). The caustic flow dominates until the flow ceases to be visible to the detector during parts of the year, resulting in a sharp drop in $T$. For the case of $\xi = 0.05$, the difference in $T$ between June and November measurements is not so large. At low energies, $T$ is mostly due to the Maxwellian, and is largest in November. With increase in recoil energy, the modulation amplitude decreases, reversing phase at $\sim 12.5$ keV. The peak now occurs in June. The next change occurs at $Q \sim 22$ keV, with $T$ measured in November being larger than in June, owing to the absence of the caustic flow near June. The phase changes again at $Q \sim 29$ keV, due to the total absence of the caustic flow. For $Q > 29$ keV, the modulation is consistent with a Maxwellian halo, with a maximum measured in June.

\begin{figure}[!ht]
\begin{center}
\scalebox{0.5}{\includegraphics{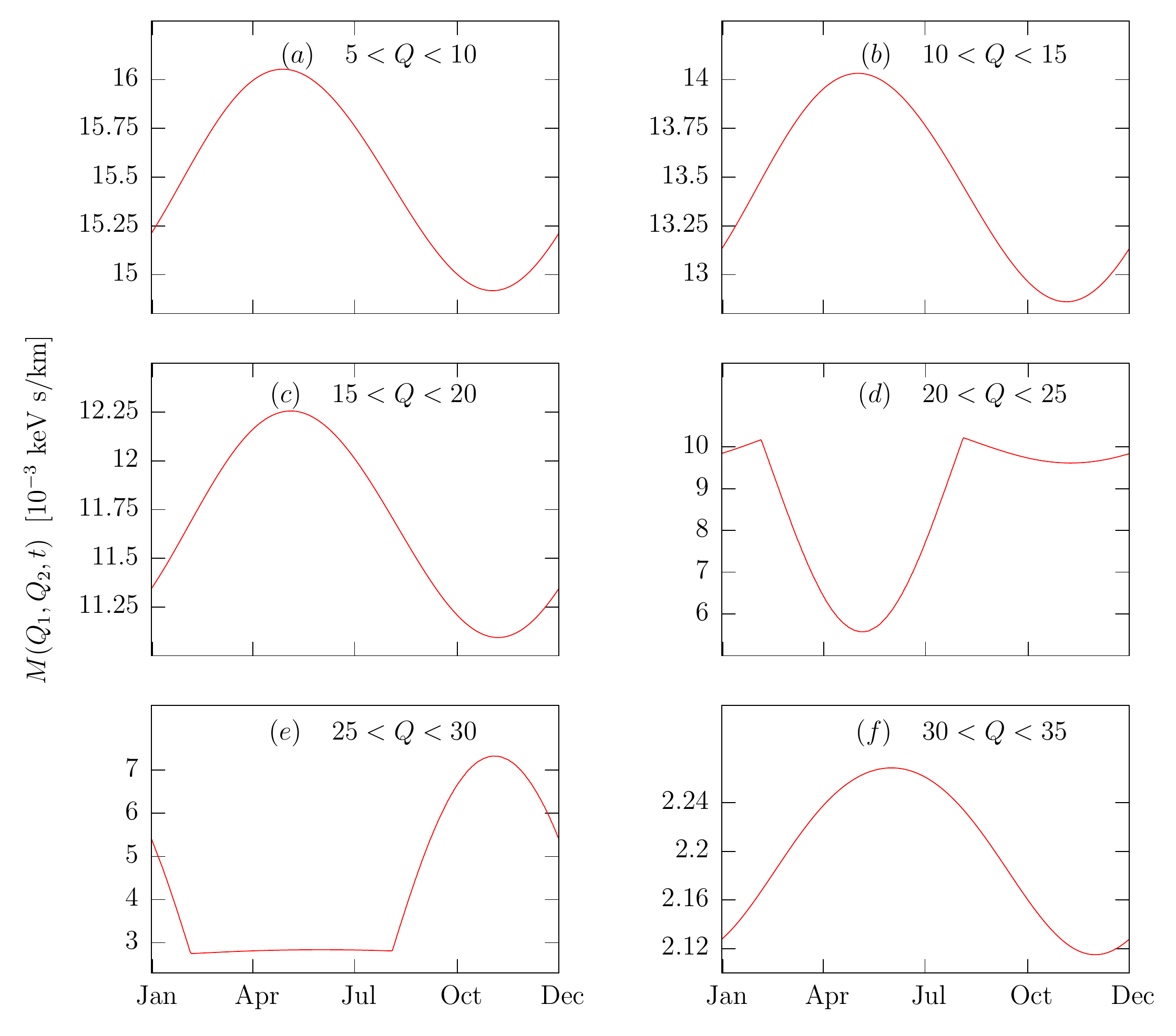}}
\end{center}
\caption{  Annual modulation in $M(Q_1,Q_2,t)$ for the caustic flow (a). $Q$ is in keV. Case $\xi = 0.5$.     \label{fig5} }
\end{figure}

Fig. \ref{fig5} shows the annual modulation in $M(Q_1,Q_2,t)$ as a function of the time of year, for the case $\xi = 0.5$ (caustic flow contributes 50\% to the local dark matter density). The six panels show different recoil energy ranges, with a bin size of 5 keV. Panels (a), (b), and (c) are qualitatively identical, and the caustic flow is dominant, with the recoil rate being largest in June, and smallest in November. The caustic flow is not seen during part of the year in Panel (d), resulting in the steep minimum around June. The modulation is about 25\%. In panel (e), much of the flow is absent around June, leading to a near $\sim 50\%$ modulation in recoil energy. The flow is completely invisible in panel (f), and the modulation of $\sim 3\%$ is due to the Maxwellian distribution.

We see from panel (a) that $M$ peaks in June and is lowest in November, in the $5-10$ keV energy band. If one were to incorrectly assume a Maxwellian distribution with no cold flow contribution, panel (a) would imply an upper limit for the WIMP mass of about $m_\chi = 30$ GeV (for the assumed $A = 73$), far below the value of $m_\chi = 70$ GeV used in this example. Thus, the phase of the annual modulation constrains the WIMP mass \cite{mass_phase} only when the distribution of particle velocities in the solar neighborhood is known. One may however use the results of two different experiments to constrain the mass without assuming a form for the velocity distribution \cite{drees1, drees2}.

\begin{figure}[!h]
\begin{center}
\scalebox{0.5}{\includegraphics{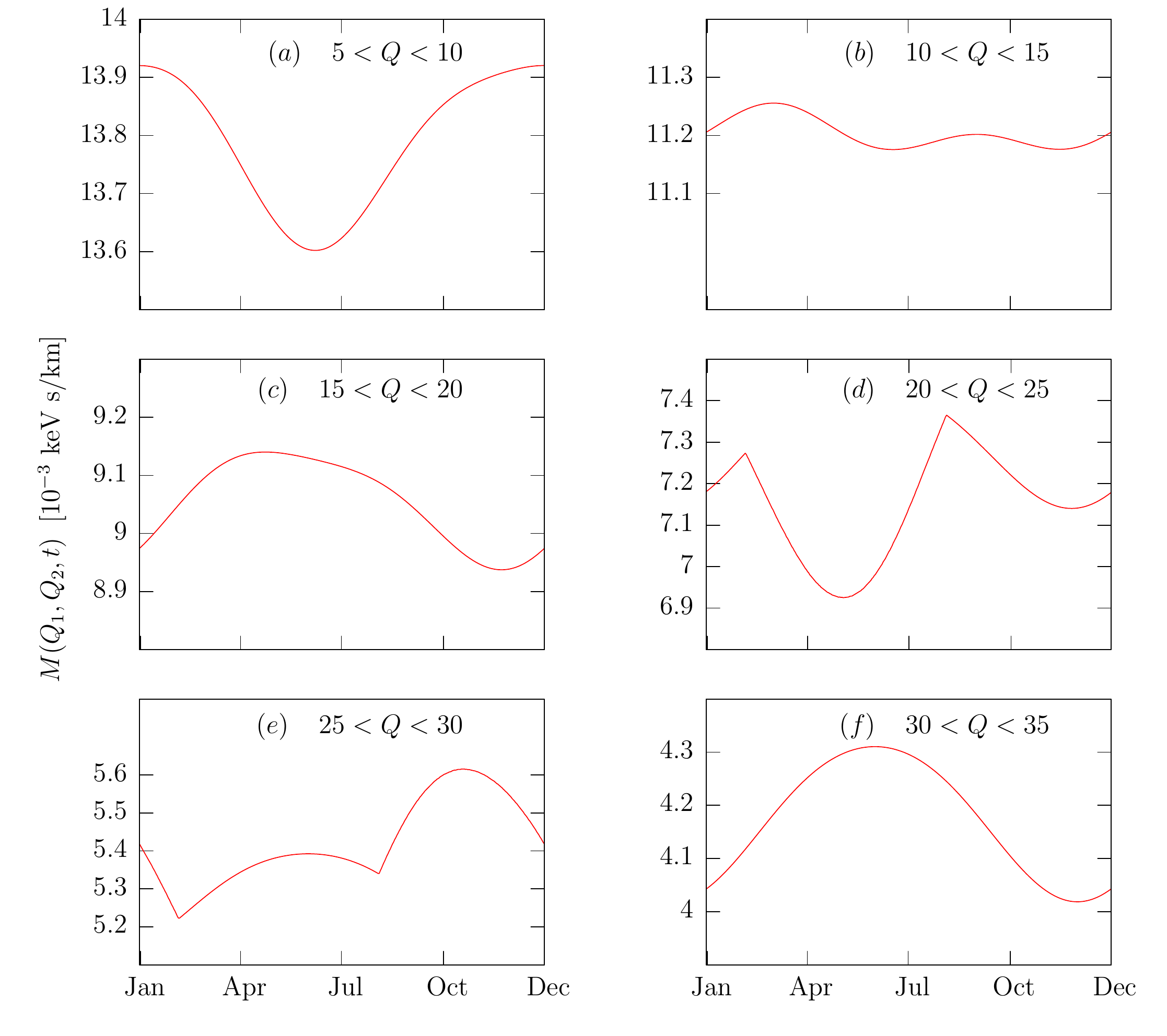}}
\end{center}
\caption{ Annual modulation in $M(Q_1,Q_2,t)$ for the caustic flow (a). $Q$ is in keV. Case $\xi = 0.05$.      \label{fig6} }
\end{figure}

Fig. \ref{fig6} shows $M(Q_1,Q_2,t)$ for the case $\xi = 0.05$ (a caustic flow contribution of 5\%). In the $5-10$ keV energy range, $M$ shows a minimum in June, and a maximum in November due to the dominant contribution of the Maxwellian halo. Phase reversal occurs in the $10-15$ keV region, and the modulation is smallest at these energies. In panel (c), $M$ peaks in June and is minimum in November. The caustic flow influences $M$ in panels (d) and (e) because at these energies, the flow is visible to the detector only during certain months of the year. As a result, even a $\sim 5\%$ contribution from the caustic flow can lead to a $\sim 5\%$ modulation effect, comparable to the modulation produced by the dominant Maxwellian component. In panel (e), the peak of $M$ occurs in November, and the modulation is far from sinusoidal. In panel (f), the caustic flow does not contribute, and the recoil rate is purely due to the Maxwellian distribution.

\section{Conclusions}

We have discussed the formation of dark matter caustics and provided possible observational evidence for the existence of outer and inner caustics. The self-similar infall model predicts the locations of the inner caustics, and their approximate densities. According to this model and the observations, the earth may be located near the fifth inner caustic. One may hope to detect the nearby caustic by means of the annual modulation expected in the recoil rate of a dark matter detector.

We discussed the annual modulation effect and derived expressions for the expected recoil rate. We considered the case of a Maxwellian velocity distribution modified by a cold flow due to the nearby caustic in the self-similar infall model.  Fig. 3 shows the velocities of the two flows that contribute to the formation of the nearest caustic. Fig. 4 shows the mean inverse speed $T(Q,t)$ expected for a cold flow, for a Maxwellian distribution, and for a combination of the two. We then computed the recoil rate integrated over energies in 5 keV bins. We considered two cases: (i) a 50\% contribution due to the caustic flow (Fig. 5)  and (ii) a 5\% contribution (Fig 6). We showed that the phase of the annual modulation at low energies can be used to constrain the WIMP mass only if the velocity distribution of dark matter particles in the solar neighborhood is known. We also showed that even a small contribution by a caustic flow can significantly alter the modulation of the recoil rate, at energies near the threshold energy of the caustic flow. Annual modulation is thus an excellent tool to detect a nearby dark matter caustic.

\acknowledgments{A.N. acknowledges financial support from the Bruce and Astrid  McWilliams postdoctoral fellowship.  }

\end{document}